\documentclass[aps,prb,twocolumn,preprintnumbers,amsmath,amssymb,superscriptaddress]{revtex4}%

\usepackage{graphicx}%
\usepackage{dcolumn}
\usepackage{amsmath}
\usepackage{color}
\usepackage{multirow}

\begin{document}

\title{Self-consistent two-gap  approach in studying multi-band superconductivity in  NdFeAsO$_{0.65}$F$_{0.35}$}

\author{Ritu Gupta}
 \affiliation{Laboratory for Muon Spin Spectroscopy, Paul Scherrer Institute, CH-5232 Villigen PSI, Switzerland}

\author{Alexander Maisuradze}
 \affiliation{Laboratory for Muon Spin Spectroscopy, Paul Scherrer Institute, CH-5232 Villigen PSI, Switzerland}
 \affiliation{Department of Physics, Tbilisi State University, Chavchavadze 3, GE-0128 Tbilisi, Georgia}

\author{Nikolai D. Zhigadlo}
 \affiliation{Department of Chemistry and Biochemistry, University of Bern, CH-3012 Bern, Switzerland }
 \affiliation{Laboratory for Solid State Physics, ETH Zurich, CH-8093 Zurich, Switzerland}
 \affiliation{CrystMat Company, CH-8046 Zurich, Switzerland}

\author{Hubertus Luetkens}
 \affiliation{Laboratory for Muon Spin Spectroscopy, Paul Scherrer Institute, CH-5232 Villigen PSI, Switzerland}

\author{Alex Amato}
 \affiliation{Laboratory for Muon Spin Spectroscopy, Paul Scherrer Institute, CH-5232 Villigen PSI, Switzerland}

\author{Rustem Khasanov}
 \email{rustem.khasanov@psi.ch}
 \affiliation{Laboratory for Muon Spin Spectroscopy, Paul Scherrer Institute, CH-5232 Villigen PSI, Switzerland}

\date{\today}

\begin{abstract}
High-quality single crystals of NdFeAsO$_{0.65}$F$_{0.35}$ (the superconducting transition temperature $T_{\rm c}\simeq 30.6$~K) were studied in zero-field (ZF) and transverse-field (TF) muon-spin rotation/relaxation ($\mu$SR) experiments. An upturn in muon-spin depolarization rate at $T\lesssim 3$~K was observed in ZF-$\mu$SR measurements and it was associated with the onset of ordering of Nd electronic moments. Measurements of the magnetic field penetration depth ($\lambda$) were performed in the TF geometry. By applying the external magnetic field $B_{\rm ex}$ parallel to the crystallographic $c$-axis ($B_{\rm ex}\| c$) and parallel to the $ab$-plane ($B_{\rm ex}\| ab$), the temperature dependencies of the in-plane component ($\lambda_{ab}^{-2}$) and the combination of the in-plane and the out of plane components  ($\lambda_{ab,c}^{-2}$) of the  superfluid density were determined, respectively. The out-of-plane superfluid density component ($\lambda_{c}^{-2}$) was further obtained by combining the results of $B_{\rm ex} \| c$ and $B_{\rm ex} \| {ab}$ set of experiments. The temperature dependencies of $\lambda_{ab}^{-2}$, $\lambda_{ab,c}^{-2}$, and $\lambda_{c}^{-2}$ were analyzed within the framework of a self-consistent two-gap model despite of using the traditional $\alpha$-model.  Interband coupling was taken into account, instead of assuming it to be zero as it stated in the $\alpha$-model.  A relatively small value of the interband coupling constant $\Lambda_{12} \simeq 0.01$ was obtained, thus indicating that the energy bands in NdFeAsO$_{0.65}$F$_{0.35}$ are only weakly coupled. In spite of their small magnitude, the coupling between the bands leads to the single value of the superconducting transition temperature $T_{\rm c}$. 
The penetration depth anisotropy $\gamma_{\lambda}=\lambda_c/\lambda_{ab}$ was found to increase upon cooling, consistent with most of Fe-based superconductors, and their behavior is attributed to the multi-band nature of superconductivity in NdFeAsO$_{0.65}$F$_{0.35}$.
\end{abstract}

\maketitle

\section{INTRODUCTION}

Iron-based superconductors (IBS's) remain  a subject of intensive research due to a comparable large value of the transition temperature $T_{\rm c}$.  It reaches up to 55 K for the $R$FeAsO$_{1-x}$F$_x$ IBS family ($R$ corresponds to the lanthanides La, Sm, Ce, Nd, Pr, and Gd),\cite{Ren_MRI_2008, Yang_SST_2008, Zhi_ChPL_2008, Chen_Nat_2008, Chen_PRL_2008}  and approaches $T_{\rm c}\simeq 100$~K in a single layer of FeSe on the SrTiO$_3$ substrate.\cite{Ge_NatMat_2015}  Emergence of superconductivity at such  high temperatures raises a puzzling question about the gap symmetry, which can further determine the pairing mechanism for the superconducting state.

The superconductivity in IBS's appears in close proximity to the magnetism offered by $d$-orbitals of Fe, hence one can expect the unconventional nature of the superconducting state. The electronic band structure calculations manifest that superconductivity in IBS's originates from multiple disconnected Fermi surface sheets derived from Fe $d$-orbitals, thus reflecting the possibility of a complex nature of the superconducting gap structure. \cite{Singh_PRL_2008, Yin_PRL_2008} There are already different scenarios proposed for the gap structure in IBS's including two-gap, $s$-wave, $d$-wave, isotropic, anisotropic, and surprisingly, $p$-type wave symmetry of the superconducting order parameter.\cite{Daghero_PRB_2008, Malone_PRB_2009, Kuzmicheva_PRB_2017, Matano_PRL_2008, Kuroki_PRL_2008, Patrick_PRB_2008, Evtushinsky_NJP_2009, Borisenko_NatPhys_2016, Charnuka_SciRep_2015_2, Khasanov_PRB_2018} Even after several years of discovery of IBS's, an unified picture of the gap structure is not reached, contradicting the case of cuprate high-temperature superconductors, where almost all superconducting families represent a nodal pairing state (see {\it e.g.} Ref.~\onlinecite{Tsuei_RMP_2000} and references therein). In the context of conflicting results, there is still a need of comprehensive tools to understand the gap symmetry of IBS's. The magnetic penetration depth and its anisotropy carry important information about the low lying quasiparticles and hence can shed light on the gap structure of IBS's.

This paper presents a detailed muon-spin rotation/relaxation ($\mu$SR) investigations of high quality single crystals of  NdFeAsO$_{0.65}$F$_{0.35}$  grown with high pressure and high temperature cubic anvil technique. Very few investigations were carried out in the direction of exploring the symmetry of order parameter for NdFeAsO$_{1-x}$F$_x$ (Nd-1111). As an example, a single gap without nodes at the $\Gamma$ hole pocket was revealed through angle resolved photoemission spectroscopy (ARPES),\cite{Kondo_PRL_2008} a nodal type gap structure was concluded through the linear behavior of the lower critical field $B_{\rm c1}$ at low temperatures.\cite{Wang_JPCM_2009} A multi-band nature of superconductivity seems to be a more generic feature for Nd-1111 as most measurements point towards a two superconducting gaps without nodes as, {\it e.g.}, the magnetic penetration depth measured through Tunnel Diode Resonator (TDR) technique,\cite{Martin_PRL_2009} ARPES,\cite{Charnuka_SciRep_2015} the point contact Andreev reflection spectroscopy,\cite{Samuley_SST_2009, Kuzmicheva_PRB_2019} conductance,\cite{Miyakawa_JSNM_2010}  and $B_{\rm c1}$ measurements.\cite{Adamski_PRB_2018} In most cases, however, the analysis of the multiple gap behavior was performed within the framework of a phenomenological
$\alpha$-model,\cite{Bouquet_EPL_2001, Carrington_PhysC_2003, Guritanu_PRB_2004, Prozorov_SST_2006,
Khasanov_PRL_2007, Khasanov_JSNM_2008, Khasanov_PRL_2009, Khasanov_PRL_2009_2, Khasanov_PRB_2014, Khasanov_PRB_2019} which assumes zero coupling between the energy bands. In fact, the zero-coupling requires that the temperature dependencies of the energy gaps, as well as the values of the
superconducting transition temperatures,  can not be identical and should vary from one to another energy band. Speaking in a broader way, there is a clear need of different set of data and an analysis taking into account the coupling between the bands. In the present paper we approached to a so-called self-consistent model,\cite{ Bussmann-Holder_EPB_2004, Kogan_PRB_2009, Bussmann-Holder_Arxiv_2009, Khasanov_PRL_2010} and used it in order to analyze the  magnetic penetration depth data obtained  in the  TF-$\mu$SR experiment  on high-quality  NdFeAsO$_{0.65}$F$_{0.35}$ single crystalline samples. Within our analysis, the energy bands with two different superconducting order parameters were assumed to be coupled and the gap equations were solved self-consistently by considering the presence of the 'interband' and 'intraband' coupling strengths.

The paper is organized as follows: In Sec.~\ref{sec:Experiental-Techniques} the sample preparation procedure, the results of magnetization measurements and the details of $\mu$SR experiments are briefly discussed. The experimental results obtained in zero-field (ZF) and transverse-field (TF) $\mu$SR experiments are described in Sec.~\ref{sec:Experimental_Results}: the subsection \ref{sec:ZF-experiments} comprises studies of the magnetic response of  NdFeAsO$_{0.65}$F$_{0.35}$, and the subsection \ref{sec:TF-experiments}
describes the results of the field-shift experiments, as well as the measurements of the  temperature dependencies of the magnetic field penetration depth. The self-consistent two-gap model and the temperature evolution of the penetration depth anisotropy
are presented in Sec.~\ref{sec:Discussions}. The conclusions follow in Sec.~\ref{sec:Conclusions}.

\section{Experimental techniques} \label{sec:Experiental-Techniques}

\subsection{Sample Preparation}

Bulk single crystals of NdFeAsO$_{1-x}$F$_x$ with nominal fluorine content $x = 0.35$ were grown at $\simeq3$~GPa and $\simeq 1450^{\rm ~o}$C from NaAs/KAs flux
by using the cubic anvil high-pressure and high-temperature technique. The detailed description of the sample preparation procedure is given in Ref. \onlinecite{Zhigadlo_PRB_2012}. The individual crystals obtained after the sample grow had a typical size of approximately 0.5x0.5x0.03~mm$^{3}$.

\subsection{Magnetization measurements}

\begin{figure}[htb]
\centering
\includegraphics[width=0.85\linewidth]{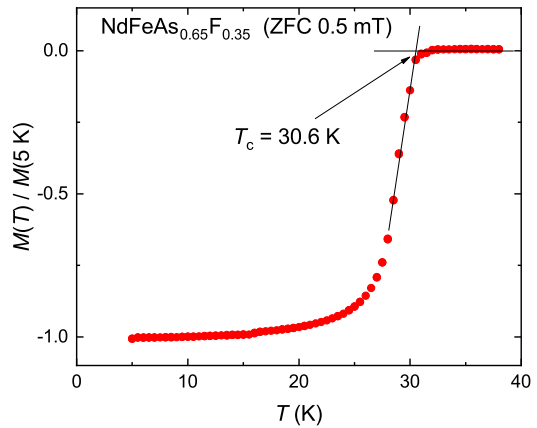}
\caption{\label{fig:magnetization} The zero-field cooled magnetization curve [$M(T)$ curve] measured on about thirty NdFeAsO$_{0.65}$F$_{0.35}$ single crystals,
which were further used in $\mu$SR experiments. The external magnetic field $B_{\rm ex} = 0.5$~mT was applied along the $ab$-plane of the crystals. The $M(T)$ data were normalized to their $M(T={\rm 5~K})$ value. The superconducting transition temperature $T_{\rm c}\simeq 30.6$~K was determined from the cross point of the two lines extrapolated from the high temperature normal state and the low temperature superconducting state, respectively.}
 \end{figure}

The magnetization measurements were carried out on a Quantum Design MPMS-5 system. Figure~\ref{fig:magnetization} shows the temperature variation of the normalized magnetic moment [$M(T)/M(T={\rm 5~K})$] measured simultaneously on about thirty NdFeAsO$_{0.65}$F$_{0.35}$ single crystals. These crystals were further used in $\mu$SR experiments. The external
field $B_{\rm ex}=0.5$~mT was applied parallel to the $ab$-plane of the crystals. Measurements were performed in the zero-field cooled (ZFC) mode. A sharp diamagnetic signal is seen across the superconducting transition, which confirms the bulk nature of the superconductivity. The superconducting transition temperature $T_{\rm c}\simeq 30.6$~K was determined from the cross point of the two lines extrapolated from the high temperature normal state and the low temperature superconducting state, respectively (see Fig. \ref{fig:magnetization}).

\subsection{Muon-spin rotation/relaxation experiments }
Muon-spin rotation/relaxation ($\mu$SR) measurements were carried out in a temperature range of 1.5 to 50~K at the GPS (General Purpose Surface)  ($\pi$M3 beam line) and DOLLY  ($\pi$E1 beam line) spectrometers at the Paul Scherrer Institut (PSI), Villigen, Switzerland. In this technique, 100\% spin-polarized muons are implanted uniformly through the sample volume, where they decay with the lifetime of 2.2 $\mu$s and the relevant decay positrons are detected successively. Muons act as sensitive magnetic probes. The spin of the muon precesses
in the local magnetic field $B_\mu$ with a frequency $\omega_\mu = \gamma_\mu B_\mu$ ($\gamma_\mu$ is muon gyromagnetic ratio, $\gamma_\mu/2\pi = 135.53$~MHz/T). The detailed description of $\mu$SR technique and its applications for studying the superconducting and magnetic samples can be found in Refs.~\onlinecite{Schenk_Book_1985, Cox_JPC_1987, Dalmas_JPCM_9_1997, Yaouanc_book_2011, Blundell_ConPhys_1999, Sonier_RMP_2000, Uemura_book_2015, Karl_PRB_2019}.

A specific sample holder was designed in order to perform $\mu$SR experiments on thin single crystals of NdFeAsO$_{0.65}$F$_{0.35}$. A mosaic of about 200 single crystals was sandwiched between two sheets made of several 0.125 nm thick Kapton layers.\cite{Kapton} The first few Kapton layers decelerate the muons from incoming beam and served the role of a degrader. The outgoing muons from the degrader were slow enough to stop inside the sample. The last few layers were used to stop the muons which still manage to pass through the sample. A schematic picture of the sample holder can be found in the Ref. \onlinecite{Khasanov_PRB_2016}. The data were analyzed using the
free software package MUSRFIT, Ref. \onlinecite{Suter_MuSRFit_2012}.

\section{Experimental Results} \label{sec:Experimental_Results}

\subsection{The magnetic response of NdFeAsO$_{0.65}$F$_{0.35}$: ZF-$\mu$SR experiments} \label{sec:ZF-experiments}

\begin{figure}[htb]
\centering
\includegraphics[width=1.03\linewidth]{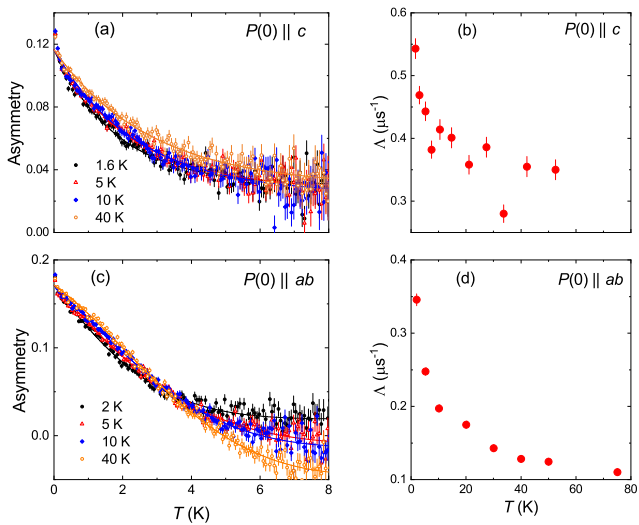}
\caption{(a) ZF-$\mu$SR time-spectra of NdFeAsO$_{0.65}$F$_{0.35}$ measured at $T=1.6$, 5, 10, and 40~K with the initial muon-spin polarization $P(0)$ applied parallel to the crystallographic $c-$axis. (b) The temperature evolution of the exponential relaxation rate $\Lambda$ obtained from the fit of Eq.~\ref{eq:ZF-TF_asymmetry} to $P(0)\| c$
set of data.  (c) and (d) -- the same as in panels (a) and (b), but for $P(0)\| ab$ set of experiments and $T=2.0$, 5, 10, and 40~K.}
 \label{fig:ZF_data}
\end{figure}

The $\mu$SR experiments in zero-field (ZF-$\mu$SR) were performed in order to study the magnetic response of the NdFeAsO$_{0.65}$F$_{0.35}$ sample. In two sets of experiments the initial muon-spin polarization $P(0)$ was applied parallel to the $c-$axis and the $ab-$plane, respectively.
Few representative muon-time spectra for $P(0)\| c$ and $P(0)\| ab$ orientations are shown in Figs. \ref{fig:ZF_data}~(a) and (c).

The experimental data were analyzed by separating the $\mu$SR response on the sample (s) and the background (bg) contributions:
\begin{equation}
 A_0P(t ) = A_{\rm s}P_{\rm s}(t ) + A_{\rm bg}P_{\rm bg}(t).
 \label{eq:ZF-TF_asymmetry}
\end{equation}
Here $A_0$ is the initial asymmetry of the muon-spin ensemble. $A_{\rm s}$ ($A_{\rm bg}$) and $P_{\rm s}(t)$ [$P_{\rm bg}(t)$] are the asymmetry and the time
evolution of the muon-spin polarization of the sample (background), respectively. The background contribution accounts for muons missing the sample and/or stopped in Kapton layers.\cite{Khasanov_PRB_2016}

In ZF-$\mu$SR experiments the sample contribution was described by assuming the presence of the nuclear and the electronic magnetic moments:
\begin{equation}
P_{\rm s}^{\rm ZF}(t) = \left[ \frac{1}{3} + \frac{2}{3}\left(1-\sigma_{\rm GKT}^2t^2\right) e^{-\sigma_{\rm GKT}^2t^2/2} \right]  e^{-\Lambda t}.
 \label{eq:ZF_polarization}
\end{equation}
Here the term within the square brackets is the Gaussian Kubo-Toyabe function with the relaxation rate $\sigma_{\rm GKT}$, which is generally used to describe the nuclear magnetic moment
contribution in ZF-$\mu$SR experiments (see, {\it e.g.}, Refs. \onlinecite{Schenk_Book_1985, Cox_JPC_1987, Dalmas_JPCM_9_1997, Yaouanc_book_2011, Blundell_ConPhys_1999,
Sonier_RMP_2000, Uemura_book_2015}, and references therein). The exponential term with the relaxation parameter $\Lambda$ represents the contribution of  randomly distributed magnetic impurities and/or disordered magnetic moments.\cite{Sonier_PRL_94, Khasanov_PRL_2009_2}

The temperature evolution of the exponential relaxation rate $\Lambda$ for $P(0)\| c$ and $P(0)\| ab$ set of experiments are presented in panels (b) and (d) of Fig.~\ref{fig:ZF_data}. During fits the Gaussian Kubo-Toyabe relaxation $\sigma_{\rm GKT}$ entering   Eq. \ref{eq:ZF_polarization} was assumed to be dependent on the orientation, but independent on temperature, respectively.
From the data presented in Figs. \ref{fig:ZF_data}~(b) and (d) two important points emerges: \\
(i) No detectable change in the relaxation rates $\Lambda$ is observed across the superconducting transition temperature, which rules out the possibility of any spontaneous magnetic field below $T_{\rm c}$. This means that the time-reversal symmetry breaking is not an immanent feature of  NdFeAsO$_{0.65}$F$_{0.35}$ studied here. \\
(ii) An increase in $\Lambda$ is seen below 3 K for both orientations, which is probably associated with the onset of ordering of Nd magnetic moments. A similar upturn was seen in measured frequency shift  [$\delta f(T)$] obtained by means of TDR technique and was explained with the ordering of the local magnetic moments of Nd  below 4 K.\cite{Martin_PRL_2009} Further evidence comes form the powder Neutron diffraction experiment, where below $\simeq1.96$~K, a long range antiferromagnetic order was apparent and it was associated to the combined magnetic ordering of Fe and Nd magnetic moments in the parent compound NdFeAsO.\cite{Qiu_PRL_2008}

\subsection{The superconducting response of NdFeAsO$_{0.65}$F$_{0.35}$: TF-$\mu$SR experiments} \label{sec:TF-experiments}

\subsubsection{The homogeneity of the superconducting state: field-shift experiments}

\begin{figure}[htb]
\centering
\includegraphics[width=1.05\linewidth]{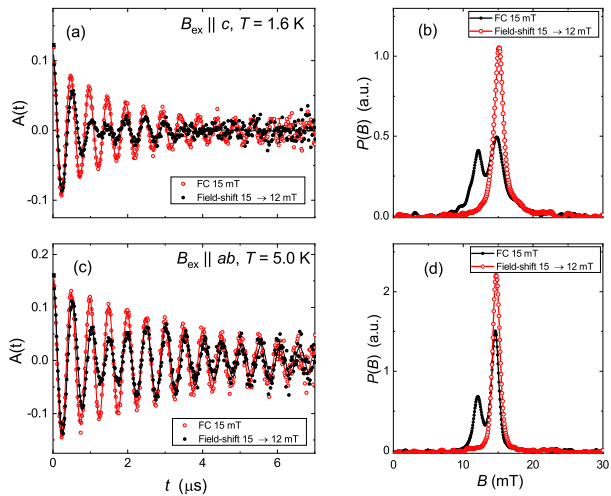}
\caption{(a) TF-$\mu$SR time-spectra  of NdFeAsO$_{0.65}$F$_{0.35}$ measured at $T=1.6$~K with the external field $B_{\rm ex}$ applied parallel to the crystallographic $c-$axis.
The TF-$\mu$SR time-spectra denoted by red points correspond to the direct cooling at $B_{\rm ex}=15$~mT from $T$ above $T_{\rm c}$ down to 1.6~K.
The black data are obtained after a subsequent field decrease to 12~mT and without changing the temperature. (b)  The Fast Fourier transform of the TF-$\mu$SR time-spectra presented
at the panel (a). (c) and (d) -- the same as in panels (a) and (b), but for $B_{\rm ex}\| ab$ and at $T=5$~K. }
\label{fig:field-shift}
\end{figure}

The homogeneity of the superconducting state and the effects of the flux-line lattice (FLL) pinning were probed by performing series of field-shift experiments in the transverse-field (TF) geometry.  The measurements were carried out with the external magnetic field $B_{\rm ex}$ applied  parallel to the $c$-axis ($B_{\rm ex} \| c$) and parallel to the $ab$-plane ($B_{\rm ex} \| ab$), respectively. The sample was initially cooled in $B_{\rm ex}\simeq 15$~mT to the desired temperature (1.6 K for $B_{\rm ex} \parallel c$ and 5~K for $B_{\rm ex} \parallel ab$) where the  first muon-time spectra were collected [red curves in Figs. \ref{fig:field-shift}~(a) and (c)]. Then, by keeping the temperature constant, the field was decreased down to 12~mT and a new 'field-shift' data sets were collected [black curves in Figs. \ref{fig:field-shift}~(a) and (c)]. The corresponding Fast Fourier transform of the TF-$\mu$SR time-spectra, which reflects the internal field distribution $P(B)$ inside the sample, are shown in panels (b) and (d) of Fig. \ref{fig:field-shift}.

The data presented in Fig. \ref{fig:field-shift}~(b) and (d) reveal that for both field orientations the main part of the signal, accounting for approximately 70\% of the total signal amplitude, remains unchanged within the experimental accuracy. Only the symmetric sharp peak follows exactly the applied field. It is attributed, therefore, to the residual background signal from muons missing the sample (see also Ref. \onlinecite{Sonier_PRL_94} where the $\mu$SR field-shift experiments were initially introduced). The field-shift experiments clearly demonstrates that for both $B_{\rm ex}\| c$ and $B_{\rm ex}\| ab$ field orientations, the flux-line lattice in NdFeAsO$_{0.65}$F$_{0.35}$ sample is strongly pinned.

\begin{figure}[htb]
\centering
\includegraphics[width=0.8\linewidth]{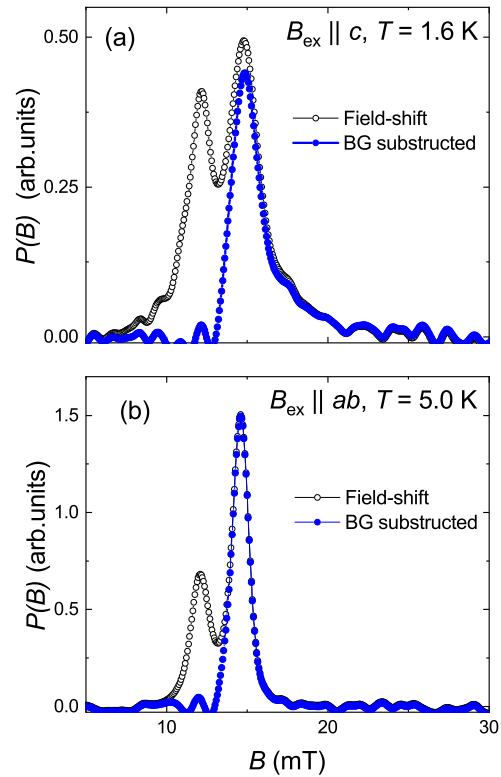}
\caption{The magnetic field distribution $P(B)$ caused by formation of the flux-line lattice in  NdFeAsO$_{0.65}$F$_{0.35}$  sample in $B_{\rm ex}\| c$ [panel (a)]
and $B_{\rm ex}\| ab$ [panel (b)] set of experiments. The blue curves are obtained by subtracting the symmetric background peak (see text for details). }
\label{fig:field-shift_substructed}
\end{figure}

The field distribution caused entirely by the flux-line lattice was further obtained by subtracting the symmetric background peak. The corresponding $P(B)'s$ are represented  in Fig.~\ref{fig:field-shift_substructed} by blue curves. It is worth noting that, for both field orientation $P(B)$ distributions possess the basic features expected for an arranged flux-line
lattice. The cutoff at low fields, the pronounced peak at the intermediate field and the long tail in the high field directions are clearly visible.

\subsubsection{Analysis of $B_{\rm ex}\| c$ and $B_{\rm ex}\| ab$ set of TF-$\mu$SR data} \label{sec:analysisi-of-TF-data}

The distribution of the internal magnetic fields $P(B)$ in the superconductor in the FLL state is uniquely determined by two characteristic lengths: the
magnetic field penetration depth $\lambda$ and the coherence length $\xi$. For an isotropic extreme type-II superconductor ($\lambda\gg\xi$) and for fields much smaller than the upper critical field $B_{\rm c2}$ ($B_{\rm ex}\ll B_{\rm c2}$) the $P(B)$ is almost independent on $\xi$ and it could be calculated from the spatial variation of the internal magnetic field $B({\bf r})$ (${\bf r}$ is the spatial coordinate).\cite{Brandt_PRB_1988, Maisuradze_JPCM_2008} In the present work the magnetic field distribution $P(B)$, measured by means of TF-$\mu$SR, was analyzed assuming $B({\bf r})$ is being described within the framework of Ginzburg-Landau approach.\cite{Brandt_PRB_1988, Maisuradze_JPCM_2008, Yaouanc_PRB_1977, Clem_JLTP_1975}

\begin{figure}[htb]
\centering
\includegraphics[width=0.85\linewidth]{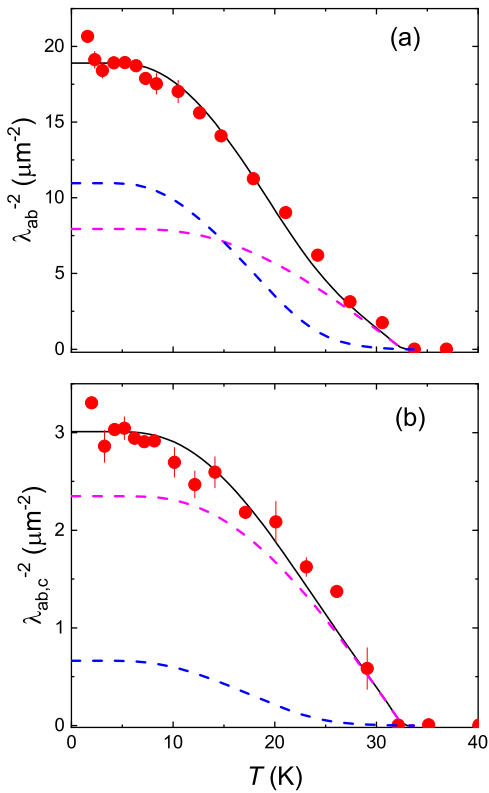}
\caption{\label{fig:lambda-ab-lambda_ac} Temperature variation of the inverse squared magnetic penetration depths:  $\lambda_{ab}^{-2}$ -- panel (a) and $\lambda_{ab,c}^{-2}$ -- panel (b). Black solid lines are theoretical curve obtained by analyzing the data within the self-consistent two-gap model. The dashed pink and blue lines are contributions of the larger ($\Delta_1$),
and the smaller ($\Delta_2$) gaps to the magnetic penetration depths, respectively. See Sec.~\ref{sec:self-consistent_model} for
details.}
\end{figure}

The spatial distribution of magnetic fields in the mixed state of a type-II superconductor is calculated via the Fourier
expansion:\cite{Brandt_PRB_1988, Maisuradze_JPCM_2008, Yaouanc_PRB_1977, Clem_JLTP_1975}
\begin{equation}
B({\bf r}) = \langle B \rangle \sum_{\bf G} \exp(-i{\bf Gr}) B_{\bf G}(\lambda\xi).
\label{eq:B(r)}
\end{equation}
Here $\big<B\big>$ is the average magnetic field inside the superconductor, ${\bf G}$ is the reciprocal vector, ${\bf r}$ represents the vector coordinate in a plane perpendicular to the applied magnetic field and $B_{\bf G}$ is the Fourier component. Within the  Ginzburg-Landau model $B_{\bf G}$ is obtained via:\cite{Yaouanc_PRB_1977}
\begin{equation}
B_{\bf G}=\frac{\phi_0}{S}(1-b^4)\frac{uK_1(u)}{\lambda^2G^2}.
\end{equation}
$\phi_0$ is the magnetic flux quantum, $S=\phi_0/\big<B\big>$ represents the area of the FLL unit cell, $b =\big<B\big>/B_{c2}$, $K_1(u)$ is the modified Bessel function, with $u^2=2\xi^2G^2(1+b^4)[1-2b(1-b)^2]$. For the hexagonal FLL, the reciprocal lattice is $G_{mn}=(2\pi/\sqrt{S})(m\cdot \sqrt[4]{3}/\sqrt{2},(n-m/2)\cdot \sqrt{2}/\sqrt[4]{3})$. $m$ and $n$ are the integer numbers.

The internal field distribution within the 'ideal' flux-line lattice was obtained as:
\begin{equation}
P_{id}(B) = \frac{\int \delta(B-B')d A(B')}{\int dA(B')}.
\label{eq:Pideal}
\end{equation}
Here $d A(B')$ is the elementary area of the FLL with a field $B'$ inside, and the integration is performed over a quarter of the flux-line lattice unit cell.\cite{Laulajainen_PRB_2006} The FLL disorder, the broadening of the TF-$\mu$SR line due to the nuclear depolarization and the contribution of the electronic moments were considering by convoluting $P_{id}(B)$ with Gaussian and Lorentzian functions.\cite{Maisuradze_JPCM_2008, Sonier_PRL_2003, Khasanov_PRL_2009_2, Khasanov_PRL_2009} Finally, the following depolarization function was fitted to the measured TF-$\mu$SR data:
\begin{equation}
P_{\rm s}^{\rm TF}(t) = e^{i\phi}e^{-\sigma_g^2 t^2/2-\Lambda t}\int P_{id}(B)e^{i\gamma_\mu Bt}dB.
 \label{eq:TF_polarization}
\end{equation}
Here $\phi$ is phase of the muon-spin ensemble, $\Lambda$ represents the relaxation rate associated with the electronic moments, and $\sigma_g$ is associated with the FLL disorder and the nuclear moments contributions, respectively. In our calculations $\Lambda$ was fixed to the values obtained in ZF-$\mu$SR experiments (see Sec.~\ref{sec:ZF-experiments}
and Fig.~\ref{fig:ZF_data}).

The results of the fit of Eq. \ref{eq:ZF-TF_asymmetry} with the sample part described by Eq. \ref{eq:TF_polarization} to the $B_{\rm ex}\| c$ and $B\| ab$ set of data are presented in Fig. \ref{fig:lambda-ab-lambda_ac}. Note that with the field applied parallel to the $c$-axis the screening current, flowing around the flux-line cores, remains within the $ab$-plane.
This means that the field distribution $P(B)$ in $B\| c$ set of experiments is  determined by the so-called in-plane component of the magnetic penetration
depth $\lambda_{ab}$ [Fig. \ref{fig:lambda-ab-lambda_ac}~(a)]. Note that in superconductors with the tetragonal layered crystal structure, as NdFeAsO$_{0.65}$F$_{0.35}$,
the $a$- and $b-$ components of the magnetic penetration depth are equal: $\lambda_{a}=\lambda_{b}$.\cite{Prozorov_SST_2006}
With the field applied parallel to the $a$($b$)-axis, the screening current flows along the $b$($a$) and $c$-axes, respectively. Consequently, in $B\| ab$ set of experiments $\lambda_{ab,c}$ is obtained [Fig. \ref{fig:lambda-ab-lambda_ac}~(b)].

\section{Discussions} \label{sec:Discussions}

\subsection{Temperature dependencies of $\lambda_{ab}^{-2}$ and $\lambda_{ab,c}^{-2}$} \label{sec:self-consistent_model}

Temperature dependencies of $\lambda_{ab}^{-2}$ and $\lambda_{ab,c}^{-2}$, as they reported in  Sec. \ref{sec:analysisi-of-TF-data}, are shown in  Figs. \ref{fig:lambda-ab-lambda_ac}~(a) and (b), respectively. Due to a possible influence caused by ordering of Nd magnetic moments
(see the discussion in Sec.~\ref{sec:ZF-experiments} and Fig.~\ref{fig:ZF_data}),
the data points below 5 K were excluded from consideration.

In order to elucidate the pairing states in NdFeAsO$_{0.65}$F$_{0.35}$, the experimental data were analyzed by means of a two-gap model,
with both gaps having an $s$-wave symmetry. Despite of considering a similar BCS type temperature dependence for both the gaps, as in phenomenological $\alpha$-model,\cite{Bouquet_EPL_2001, Carrington_PhysC_2003, Guritanu_PRB_2004, Prozorov_SST_2006,
Khasanov_PRL_2007, Khasanov_JSNM_2008, Khasanov_PRL_2009_2, Khasanov_PRB_2014, Khasanov_PRB_2019} the temperature dependencies of the two gaps ($\Delta_1$ and $\Delta_2$) were obtained through a self-consistent coupled gap equations:\cite{Bussmann-Holder_EPB_2004, Bussmann-Holder_Arxiv_2009, Khasanov_PRL_2010}
\begin{widetext}
\begin{equation}
\Delta_1 = \int_{0}^{\omega_{D1}}\frac{N_1(0)V_{11}\Delta_1}{\sqrt{E^2+\Delta_1^2}}\tanh \frac{\sqrt{E^2+\Delta_1^2}}{2k_BT}dE +
\int_{0}^{\omega_{D2}}\frac{N_2(0)V_{12}\Delta_2}{\sqrt{E^2+\Delta_2^2}}\tanh \frac{\sqrt{E^2+\Delta_2^2}}{2k_BT}dE, \nonumber
\end{equation}
\begin{equation}
\Delta_2 = \int_{0}^{\omega_{D1}}\frac{N_1(0)V_{21}\Delta_1}{\sqrt{E^2+\Delta_1^2}}\tanh \frac{\sqrt{E^2+\Delta_1^2}}{2k_BT}dE +
\int_{0}^{\omega_{D2}}\frac{N_2(0)V_{22}\Delta_2}{\sqrt{E^2+\Delta_2^2}}\tanh \frac{\sqrt{E^2+\Delta_2^2}}{2k_BT}dE.
 \label{eq:Coupled-Gaps_Full}
\end{equation}
\end{widetext}
Here, $N_1(0)$ and $N_2(0)$ are the partial density of states for each band at the Fermi level. $V_{11}$ ($V_{22}$) and $V_{12}$ ($V_{21}$)
are the intraband and the interband interaction potentials, respectively. 

A simplification of the above expressions is further done by using the notation for the coupling constant, $\Lambda_{ij}=N_j(0)V_{ij}$, which is introduced by Kogan $et$ $al.$ in Ref.~\onlinecite{Kogan_PRB_2009}. Another simplification is made by assuming similar Debye frequencies  for both the bands, $i.e.$, $\omega_{D1}=\omega_{D2}=\omega_D$. By doing this, the gap equation becomes:\cite{Bussmann-Holder_Arxiv_2009, Khasanov_PRL_2010}
\begin{widetext}
\begin{equation}
\Delta_1 = \int_{0}^{\omega_{D}}\frac{\Lambda_{11}\Delta_1}{\sqrt{E^2+\Delta_1^2}}\tanh \frac{\sqrt{E^2+\Delta_1^2}}{2k_BT}dE +
\int_{0}^{\omega_{D}}\frac{\Lambda_{12}\Delta_2}{\sqrt{E^2+\Delta_2^2}}\tanh \frac{\sqrt{E^2+\Delta_2^2}}{2k_BT}dE, \nonumber
\end{equation}
\begin{equation}
\Delta_2 = \int_{0}^{\omega_{D}}\frac{\Lambda_{12}\Delta_1}{\sqrt{E^2+\Delta_1^2}}\tanh \frac{\sqrt{E^2+\Delta_1^2}}{2k_BT}dE +
\int_{0}^{\omega_{D}}\frac{\Lambda_{22}\Delta_2}{\sqrt{E^2+\Delta_2^2}}\tanh \frac{\sqrt{E^2+\Delta_2^2}}{2k_BT}dE.
 \label{eq:Coupled-Gaps_Simple}
\end{equation}
\end{widetext}

The advantage of using the above introduced simplifications is that: \\
 (i) Within the notation of Kogan $et$ $al.$\cite{Kogan_PRB_2009}  $\Lambda_{12}=\Lambda_{21}$. \\
 (ii) The number of the free parameters, which were initially  8 in Eq. \ref{eq:Coupled-Gaps_Full} [namely: $\omega_{D1}$, $\omega_{D2}$, $N(0)_1$, $N(0)_2$, $V_{11}$, $V_{12}$, $V_{21}$, and $V_{22}$], reduces to 4 in Eq. \ref{eq:Coupled-Gaps_Simple} [namely: $\omega_D$, $\Lambda_{11}$, $\Lambda_{12}$, and $\Lambda_{22}$].\cite{Bussmann-Holder_Arxiv_2009, Khasanov_PRL_2010}

With the known temperature variation of $\Delta_1(T)$ and $\Delta_2(T)$, a rigorous analysis of $\lambda^{-2}$ is carried out by separating it into two components:\cite{Kogan_PRB_2009, Khasanov_PRL_2010}
\begin{equation}
\frac{\lambda^{-2}(T)}{\lambda^{-2}(0)}=\omega\frac{\lambda_{1}^{-2}(T)}{\lambda_1^{-2}(0)}+(1-\omega)\frac{\lambda_{2}^{-2}(T)}{\lambda_{2}^{-2}(0)}.
\end{equation}
$\omega$ is the weight factor for the larger gap $\Delta_1$ and $\lambda_{i}^{-2}(T)/\lambda_i^{-2}(0)$ is the superfluid density component of the $i-$th band.
The superfluid density component is related to the superconducting energy gap via the expression:\cite{Tinkham_book_1975}
\begin{equation}
\frac{\lambda_{i}(T)^{-2}}{\lambda_{i}(0)^{-2}}=1+2\int_{\Delta_i(T)}^{\infty}\bigg(\frac{\partial f}{\partial E}\bigg)\times \frac{E dE}{\sqrt{E^2-\Delta_i(T)^2}},
\end{equation}
where $f=[1+\exp (E/k_B T)]^{-1}$ is the Fermi distribution function.

\begin{figure}[htb]
\centering
\includegraphics[width=0.8\linewidth]{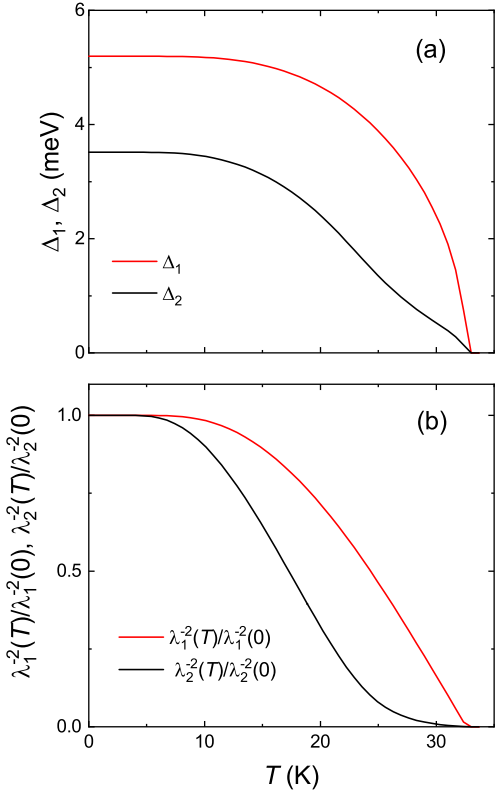}
\caption{\label{fig:gaps_lambdas} (a) Temperature evolutions of the bigger ($\Delta_1$) and the smaller ($\Delta_2$) superconducting energy gaps obtained within the framework
of the self-consistent two-gap model. (b) Temperature evolutions of the corresponding superfluid density contributions $\lambda_1^{-2}(T)/\lambda_1^{-2}(0)$ and
$\lambda_2^{-2}(T)/\lambda_2^{-2}(0)$. See Sec.~\ref{sec:self-consistent_model} for
details.  }
\end{figure}

For the analysis of the temperature evolution of the magnetic penetration depths, the literature value of the Debye frequency, $\omega_D = 37$~meV, obtained in M\"{o}ssbauer experiments,\cite{Pissas_SST_2008} was considered. The coupling constants: $\Lambda_{11}$, $\Lambda_{22}$, and $\Lambda_{12}$; the gaps: $\Delta_1(T)$, $\Delta_2(T)$
were kept identical during the analysis of $\lambda_{ab}^{-2}(T)$ and $\lambda_{ab,c}^{-2}(T)$, but the weight factor $\omega$ was varied. The common parameters obtained with the analysis  of $\lambda_{ab}^{-2}(T)$ and $\lambda_{ab,c}^{-2}(T)$ are: $\Lambda_{11} \simeq 0.368$, $\Lambda_{22} \simeq 0.315$, $\Lambda_{12} \simeq 0.01$, $\Delta_1(0) \simeq 5.2$~meV, $\Delta_2(0) \simeq 3.5$~meV, and $T_{\rm c} \simeq 33.7$~K.
The weighting factors ($\omega$) and the zero-temperature values of the inverse squared magnetic penetration depth [$\lambda^{-2}(0)$] are  0.42/0.85 and 18.9/3.0~$\mu$m$^{-2}$
for $\lambda_{ab}^{-2}(T)$ and $\lambda_{ab,c}^{-2}(T)$, respectively.

Contribution of the penetration depths corresponding to the larger gap ($\Delta_1$) and the smaller gap ($\Delta_2$) are shown in Figs. \ref{fig:lambda-ab-lambda_ac}~(a) and (b) by dashed pink  and blue lines, respectively. The solid black lines are the theorey curves obtained by means of
two-gap model as described earlier.  The temperature dependencies of the gaps [$\Delta_1(T)$ and $\Delta_2(T)$] and the corresponding superfluid density components
[$\lambda_{1}^{-2}(T)/\lambda_1^{-2}(0)$ and $\lambda_{2}^{-2}(T)/\lambda_2^{-2}(0)$] are presented in Fig.~\ref{fig:gaps_lambdas}.

From the analysis of the magnetic penetration depths based on the  self-consistent two-gap model three following important points emerge:\\
 (i) The interband coupling constant $\Lambda_{12} \simeq  0.01$ is relatively small,
 indicating the fact that the two bands are nearly decoupled. However, the value of $\Lambda_{12}$ is significant enough to assign a single $T_{\rm c}$ for each gap along both the planes.\\
 (ii) The gap to $T_{\rm c}$ ratio for the bigger gap 2$\Delta_1/k_BT_{\rm c}\simeq 3.58$ is close to the universal BCS value 3.52. For the lower gap
 2$\Delta_2/k_BT_{\rm c} = 2.41$ is determined. This indicates the weak coupling regime for both the gaps.\\
 (iii) The difference in the temperature variation of $\lambda_{ab}^{-2}$ and $\lambda_{ab,c}^{-2}$ arises because
 of much smaller contribution of larger gap to $\lambda_{ab}^{-2}$ compared to that to $\lambda_{ab,c}^{-2}$.

\subsection{Out of plane magnetic penetration depth, $\lambda_{c}^{-2}$}

This section describes the determination of the out of plane component of the magnetic penetration depth, $\lambda_{c}^{-2}(T)$, and its analysis based on the self-consistent two-gap model.

According to the London model, the inverse squared magnetic field penetration depth for the isotropic superconductor
is proportional to the superfluid density in terms of $\lambda^{-2}\propto \rho_s=n_s/m^\ast$ ($\rho_s$ is the superfluid density, $n_s$ is
the charge carrier concentration and $m^\ast$ is the effective mass of the charge carriers). For an anisotropic superconductor, as  NdFeAsO$_{0.65}$F$_{0.35}$,
the magnetic penetration depth  is also anisotropic and is determined by an effective mass tensor:\cite{Thiemann_PRB_1989}
\begin{equation}
m_{eff}=\left(\begin{array}{ccc} m_i^* & 0 & 0\\ 0 & m_j^* & 0\\ 0 & 0 & m_k^* \end{array}\right).
\end{equation}
Here, $m_i^*$ is the effective mass of charge carrier flowing along $i$-th principal axis. For a magnetic field applied along $i$-th principal axis of the effective mass tensor, the effective penetration depth is given as:\cite{Thiemann_PRB_1989}
\begin{equation}
 \lambda_{jk}^{-2}=\frac{1}{\lambda_j \lambda_k}.
 \label{eq:lambda_jk}
\end{equation}

\begin{figure}[htb]
\centering
\includegraphics[width=0.85\linewidth]{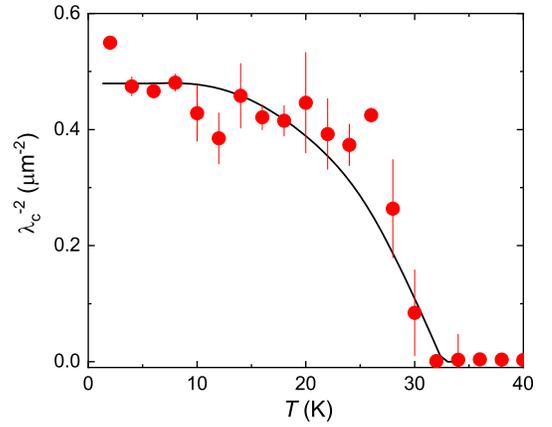}
\caption{\label{fig:lambda_c} Temperature evolution of the inverse squared out of plane magnetic penetration depth $\lambda_{c}^{-2}(T)$. The solid line is obtained from the theoretical $\lambda_{ab}^{-2}(T)$ and $\lambda_{ab,c}^{-2}(T)$ curves, as they presented in Fig.~\ref{fig:lambda-ab-lambda_ac}, by using Eq.~\ref{eq:lambda_c}.
}
\end{figure}

By using Eq.~\ref{eq:lambda_jk} the out of plane component of the magnetic penetration depth, $\lambda_c^{-2}$, was further obtained from $\lambda_{ab}^{-2}(T)$
and $\lambda_{ab,c}^{-2}(T)$ data
shown in Fig. \ref{fig:lambda-ab-lambda_ac} as:
\begin{equation}
\lambda_{c}^{-2} = \frac{\lambda_{ab,c}^{-4}}{\lambda_{ab}^{-2}}.
 \label{eq:lambda_c}
 \end{equation}

The resulting dependence of $\lambda_c^{-2}$ on temperature is shown in Fig.~\ref{fig:lambda_c}. The theoretical temperature variation of $\lambda_c^{-2}(T)$ was also obtained
from the theory curves for $\lambda_{ab}^{-2}(T)$ and $\lambda_{ab,c}^{-2}(T)$, as they described in Figs.~\ref{fig:lambda-ab-lambda_ac}~(a) and (b), and
it is represented by solid black line. It is evident that the curve obtained by means of two-gap model replicates the experimental data very well,  which indicates that the magnetic penetration depth along $c$-axis is well analyzed with two-gap $s+s$-wave model. For the zero-temperature value of the out-of plane component the value  $\lambda_c^{-2}(0)\simeq 0.48$~$\mu$m$^{-2}$ is obtained.

\subsection{Magnetic penetration depth anisotropy,  $\gamma_{\lambda}$}

\begin{figure}[htb]
\centering
\includegraphics[width=0.85\linewidth]{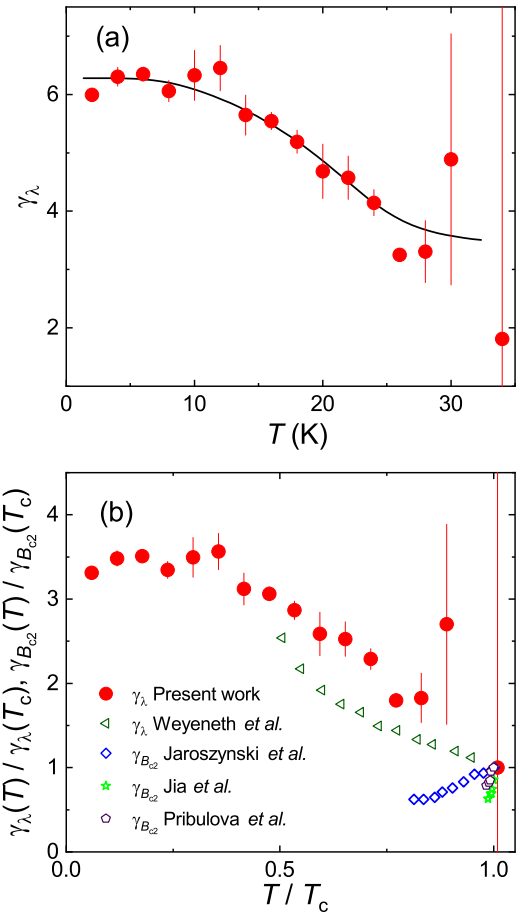}
\caption{\label{fig:gamma}  (a) Temperature dependence of the magnetic penetration depth anisotropy, $\gamma_\lambda(T)=\lambda_{c}(T)/\lambda_{ab}(T)$. The
solid curve is the result of the self-consistent two-gap model. (b) Comparison of $\gamma_\lambda(T)/\gamma_{\lambda}(T_{\rm c})$ (from the present
work and Ref. \onlinecite{Weyeneth_JSNM_2009}) with the anisotropy of the upper critical field $\gamma_{B_{\rm c2}}(T)/\gamma_{B_{\rm c2}}(T_{\rm c})$ from
Refs. \onlinecite{Jarozynski_PRB_2008, Jia_APL_2008,Pribulova_PRB_2009}. }
\end{figure}

Figure~\ref{fig:gamma}~(a) shows the temperature evolution of the magnetic penetration depth anisotropy obtained  with the experimental data presented in Fig.~\ref{fig:lambda-ab-lambda_ac} and Eqs.~\ref{eq:lambda_jk}, \ref{eq:lambda_c}:
\begin{equation}
\gamma_{\lambda} = \frac{\lambda_c}{\lambda_{ab}} = \frac{\lambda_{ab}^{-2}}{\lambda_{ab,c}^{-2}}.
\label{eq:gamma_lambda}
\end{equation}
$\gamma_{\lambda}(T)$ increases with decreasing temperature from $\gamma_{\lambda}\simeq 1.8 $ at $T=T_{\rm c}$ to $\gamma_{\lambda}\simeq 6.3 $ close to $T=0$ K. The theoretical curve obtained with two-band model is represented by the solid black line. The temperature variation of anisotropy is reproduced well with this theoretical curve, which further confirms the multi-band nature of superconductivity in the studied oxypnictide material. It  is worth to mention, that $\lambda_{ab}^{-2}(T)$ and $\lambda_{ab,c}^{-2}(T)$, obtained within the present study, were measured on a mosaic of about 200  NdFeAsO$_{0.65}$F$_{0.35}$ single crystalline samples. For such a big number of simultaneously measured crystals a certain misalignment will definitively take place. Consequently, our results put a lower limit on the determination of  $\gamma_{\lambda}$.

Figure~\ref{fig:gamma}~(b) compares  $\gamma_{\lambda}$ obtained  in the present study with that measured by means of torque magnetometry by Weyeneth {\it et al.} in Ref.~\onlinecite{Weyeneth_JSNM_2009}. In both cases $\lambda_\lambda$ increases with decreasing $T$.
A similar qualitative behavior of $\gamma_{\lambda}(T)$ was observed in Sm- and Nd-1111 system by means of torque magnetometry;\cite{Weyeneth_JSNM_2009, Weyeneth_JSNM_2009_2} in Ba(Fe$_{1-x}$Co$_x$)$_2$As$_2$ by means of TDR;\cite{Prozorov_PhysC_2008} in Ba$_{1-x}$K$_x$Fe$_2$As$_2$,\cite{Khasanov_PRL_2009}  SrFe$_{1.75}$Co$_{0.25}$As$_2$,\cite{Khasanov_PRL_2009_2} FeSe$_{0.5}$Te$_{0.5}$,\cite{Bendele_PRB_2010}   CaKFe$_4$As$_4$,\cite{Khasanov_PRB_2019} by means of $\mu$SR {\it etc.} In all these works the pronounced temperature dependence of $\gamma_{\lambda}$ was attributed to the multiple gap nature of superconductivity.

As a further step, $\gamma_{\lambda}$ is compared with the anisotropy of the  upper critical field $\gamma_{B_{c2}}$ for  NdFeAsO$_{1-x}$F$_x$, as obtained form resistivity \cite{Jarozynski_PRB_2008, Jia_APL_2008} and specific heat measurements.\cite{Pribulova_PRB_2009} According to the phenomenological Ginzburg-Landau theory, these two anisotropies should be equal for a single gap superconductor:\cite{Kogan_PRB_1981, Tinkham_book_1975}
\begin{equation}
\gamma_\lambda=\frac{\lambda_c}{\lambda_{ab}}=\sqrt{\frac{m_c^*}{m_{ab}^*}}=\gamma_{B_{\rm c2}}=\frac{B_{\rm c2}^{\parallel ab}}{B_{\rm c2}^{\parallel c}}=\frac{\xi_{ab}}{\xi_c}.
\end{equation}
Figure~\ref{fig:gamma}~(b) implies that the two anisotropies show opposite trends with temperature and violates the Ginzburg-Landau theory. This situation is reminiscent of the well known two-gap superconductor MgB$_2$, despite the reversed slope for both the anisotropies.\cite{Angst_PRL_2002, Fletcher_PRL_2005}

\section{Conclusions} \label{sec:Conclusions}

To conclude, the magnetic and superconducting properties of NdFeAsO$_{0.65}$F$_{0.35}$ single crystalline samples were studied by means of muon-spin rotation/relaxation technique. The results of the studies are summarised as follows: \\
(i) No changes in the relaxation rate were observed in ZF-$\mu$SR spectra across the superconducting transition, thus ruling out the possibility of any spontaneous magnetic field  below $T_{\rm c}$.\\
(ii) An upturn in exponential muon-spin depolarization rate at $T\lesssim 3$~K, is detected in ZF-$\mu$SR measurements. It is most probably  associated with the onset of ordering of Nd electronic moments.\\
(iii) Measurements of the magnetic field penetration depth ($\lambda$) were performed in the TF geometry. By applying the external magnetic field $B_{\rm ex}$ parallel to the crystallographic $c$-axis and parallel to the $ab$-plane, the temperature dependencies of the in-plane component $\lambda_{ab}^{-2}$ and the combination of the in-plane and the out of plane components  $\lambda_{ab,c}^{-2}$ of the  superfluid density were determined, respectively. The out-of-plane component $\lambda_{c}^{-2}$ was further obtained by combining the results of $B_{\rm ex} \| c$ and $B_{\rm ex} \| {ab}$ set of experiments. \\
(iv) The temperature dependencies of $\lambda_{ab}^{-2}$, $\lambda_{ab,c}^{-2}$, and $\lambda_{c}^{-2}$ were analyzed within the framework of a self-consistent two-gap model despite of using the traditional $\alpha$-model.  Interband coupling is taken into account instead of assuming it to be zero as is assumed in the $\alpha$-model. The values of intraband and interband coupling constats were determined to be: $\Lambda_{11} \simeq 0.368$, $\Lambda_{22} \simeq 0.315$, and $\Lambda_{12}\simeq 0.01$.   A relatively small value of the interband coupling constant $\Lambda_{12}$  indicates that the energy bands in NdFeAsO$_{0.65}$F$_{0.35}$ are nearly decoupled. \\
(v) The zero-temperature values of the inverse squared magnetic penetration depth and the superconducting energy gaps were estimated to be: $\lambda_{ab}^{-2}(0)\simeq 18.9$~$\mu$m$^{-2}$,  $\lambda_c^{-2}(0)\simeq 0.48$~$\mu$m$^{-2}$, $\Delta_1(0) \simeq 5.2$~meV, and $\Delta_2(0) \simeq 3.5$~meV, respectively.\\
(vi) The magnetic penetration depth anisotropy, $\gamma_\lambda=\lambda_{c}/\lambda_{ab}$ increases from $\gamma_\lambda \simeq 1.8$  at $T=T_{\rm c}$ to $\gamma_{\lambda}\simeq 6.3 $ close to $T=0$ K, while the upper critical field anisotropy $\gamma_{B_{\rm c2}}$ demonstrates the opposite temperature behaviour. This experimental situation is similar to MgB$_2$, a well known two-gap superconductor, which further provides a strong evidence for multiple band superconductivity in the studied NdFeAsO$_{0.65}$F$_{0.35}$ compound.

\section{Acknowledgments}
This work was performed at Swiss Muon Source (S$\mu$S), Paul Scherrer Institute (PSI, Switzerland).
RK, AM, and NDZ thank Bertram Batlogg for fruitful discussions on the early stage of this study.
The work of RG was supported
by the Swiss National Science Foundation (SNF-Grant No. 200021-175935).

\end{document}